\documentclass{aastex631}
\usepackage{subfigure}
\usepackage{natbib}

\begin{document}

\title{LTD064402+245919, compact binary or subgiant with red star eclipsing binary?}
\author{Jincheng Guo}
\altaffiliation{Department of Scientific research, Beijing Planetarium, Xizhimenwai Road, Beijing 100044, China; }
\author{Cheng Liu}
\altaffiliation{Department of Scientific research, Beijing Planetarium, Xizhimenwai Road, Beijing 100044, China; }
\email{$Email$: Andrewbooksatnaoc@gmail.com (Jincheng Guo)}

\begin{abstract}
Recently, a single-line spectroscopic binary, LTD064402+245919, has been discovered by Yang et al. Using data from LAMOST and ZTF, the unseen companion is estimated to have a mass of 1-3\,$M_{\odot}$, orbiting a subgiant with orbital period of 14.50 days, making it a good compact binary candidate without X-ray emission. However, new light curves from ZTF and ASAS-SN, have shown the depth of one dip increases towards a bluer wavelength, indicating LTD064402+245919 is more likely to be a subgiant with a red star. Using both Wilson-Devinney code and Phoebe, the derived $T_{ eff}$ of secondary is about 3400 \,K, corresponding to a red M2/3 star. Additionally, the 20\% error of parallax from Gaia is large. The mass of subgiant will be 1.28\,$M_{\odot}$ instead of 2.77\,$M_{\odot}$, if the refined distance of 5.0\,kpc is used. Nevertheless, new multi-colour photometry are warranted for the final confirmation of binary properties.

\end{abstract}


 \section{Introduction}
Compact objects like neutron stars (NSs) and black holes (BHs) are rare objects of great importance. Currently the most effective way to identify them is through X-ray binary, when accretion causing the binary to be detectable as an X-ray transient \citep{McClintock2003, Casares2014}. Meanwhile, new BH and NS binary candidates without X-ray emission have been discovered via spectroscopic surveys like SDSS and LAMOST \citep{Thompson2019,Liu2019}. 

More recently, new discovery of a subgiant with a $1-3\,M_{\odot}$ unseen companion, LTD064402+245919 (hereafter LTD0644),  has been reported by \cite{Yang2021}. Based on inverse parallax from Gaia DR2 and spectra from LAMOST Time-domain survey \citep[TD;][]{Wang2021}, the observed primary (visible component) is estimated to be a subgiant with  D=$6.7^{+1.0}_{-0.8}$\,kpc, $T_{eff}=4500\pm200$\,K, log\,$g$=2.50$\pm$0.25\,dex, [Fe/H]=-0.54$\pm$0.18\,dex, and mass=2.77$\pm0.68\,M_{\odot}$. By combining $r$ band light curve (LC) from ZTF \citep{Masci2019} and radial velocity (RV) curve from LAMOST, \citet{Yang2021} used PHOEBE \citep{Prsa2005, Conroy2020} to model the mass of secondary M=$2.02\pm0.49\,M_{\odot}$ with inclination angle $i\approx47^{\circ}$, assuming it is a main-sequence star, and the secondary mass will be above $1\,M_{\odot}$ when compact binary is assumed. 

\begin{figure*}[!htbp]
\center
\includegraphics[angle=0,width=0.8\textwidth]{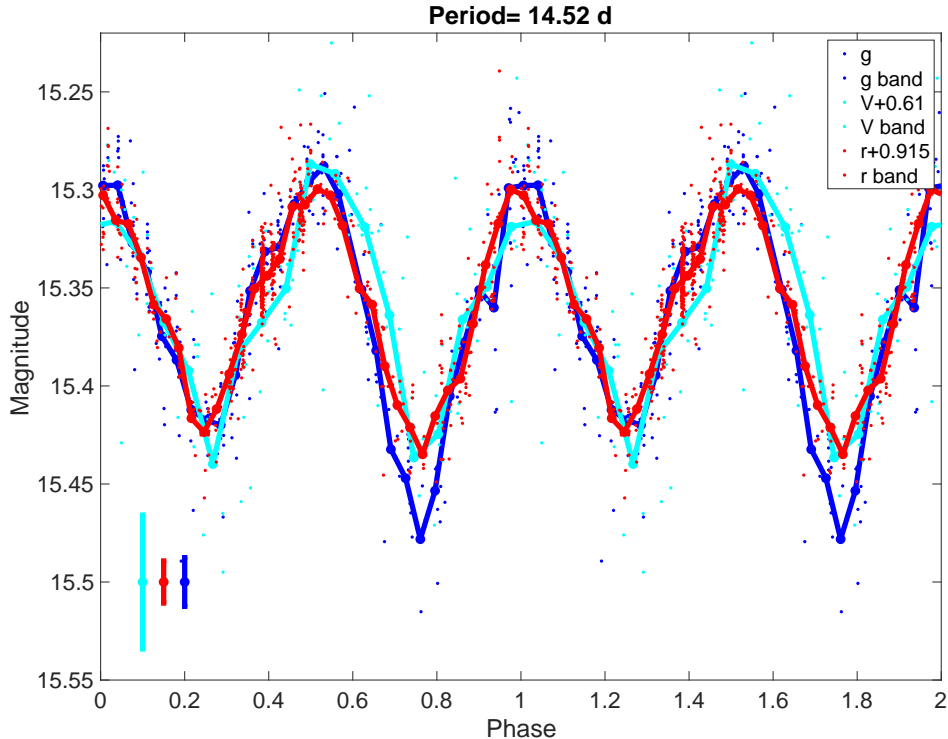}
\caption{Folded LCs from $g$, V and $r$ band are plotted in blue, cyan and red, respectively. Lines are re-binned observations. Folded LCs of $r$ and V band are shifted to the same level as $g$ band ($r$+0.915\,mag, V+0.61\,mag). Median photometric errors are shown on the lower left.}
\label{fig1}
\end{figure*}

\section{Data and Analysis}
LTD0644 with R.A. and decl. of $101.0089^{\circ}, 24.9887^{\circ}$ has been observed by ZTF, Pan-STARRS \citep{Chambers2016} and ASAS-SN \citep{Kochanek2017}. By the time of April 2022, there are 283 and 731 good observations (catflag=0) recorded by ZTF DR10 in $g$ and $r$ bands, respectively \footnote{\url{https://irsa.ipac.caltech.edu/cgi-bin/Gator/nph-dd}}. ASAS-SN has observed 118 data points in V band, while 14, 6, 17, 17 and 21 data points are recorded by Pan-STARRS in $g, r, i, z $ and $y$ bands, respectively. These Pan-STARRS data are limited in number, but do provide colour information. 

An updated orbital period of 14.52 d is obtained by applying Lomb-Scargle \citep{Lomb1976,Scargle1982} to ZTF LC in $r$ band, which has the most observed data points. By folding the LCs from ZTF and ASAS-SN with 14.52 d, then shifting $r$ and V band to the same level as $g$ band, an important feature in the LC has been noticed (Figure. \ref{fig1}). The depth of one dip, i.e. near phase 0.8, increases towards a bluer wavelength, while the depth of the other dip stays almost the same. This is a typical feature of an eclipsing binary consisting of two normal stars. It can be explained by a dimmer, red star transiting a hotter primary. When the primary is being transited by a dimmer, red star, the secondary will contribute less brightness to the blue band LC during eclipse, compared to the red band eclipse. In contrast, when the primary is eclipsing secondary, colour change is mostly caused by colour difference in subgiant itself, which is much smaller than the colour change during primary eclipse. Therefore, the compact binary scenario is less favoured, as the brightness contributed from NS and BH is negligible or none and white dwarf is blue. In other words, the LC variation should mostly come from the rotation of a distorted companion star. Yet, the rotation of a deformed star alone is not able to produce such a colour difference in the depth. 

By using Wilson-Devinney code \citep[WD,][]{Wilson1971}, we have fitted LCs from ZTF in $g$ and $r$ band, together with RV curve from LAMOST TD. The same $T_{eff}$ of 4500\,K, log\,$g$ of 2.5 and [Fe/H] of -0.5 are used as constraint parameters. The secondary $T_{eff}$=3411\,K, $i=78^{\circ}$ and mass ratio $q$=0.72 are derived. Same input data and constraint parameters are also adopted for Phoebe. The derived $T_{eff}$ of secondary is 3456\,K, similar with WD code estimation, consistent with a red M2/3 type star \citep{Cox2000}.

In addition, the inverse parallax distance of 6.7\,kpc adopted by \cite{Yang2021} maybe debatable, since a refined distance of 5.0\,kpc has been provided by \cite{Bailer-Jones2021}, using data from Gaia EDR3. This distance would derive the primary mass M=1.28\,$M_{\odot}$ instead of 2.77\,$M_{\odot}$. The undetected companion mass will be $0.93\,M_{\odot}$ instead of $2.02\,M_{\odot}$, if $q$=0.73 from \cite{Yang2021} is used. One should note that even 5.0\,kpc may be overestimated. Thus, the rough mass estimation here is only to make a point that distance is crucial in mass determination. It is likely that future release of Gaia parallax will provide more accurate and consistent distance for LTD0644.

\section{Summary}
In this paper, we report our new findings from multi-band LCs regarding the newly discovered LAMOST binary LTD0644. Based on the fact that the depth of one dip increases towards bluer band, the binary system LTD0644 with 14.52 d orbital period is more likely to be a subgaint with a red star in an eclipsing binary system. According to the WD code and Phoebe modelling of LCs in $g$, $r$ band and RV curve, $i$=78$^{\circ}$ and $q$=0.72 are estimated. The secondary $T_{eff}\approx$3400\,K is derived, consistent with a red M2/3. However, more accurate multi-colour LCs are needed in order to derive reliable results. It is noted that in hunting for noninteracting compact binaries, multi-colour information is crucial to rule out the normal binary scenarios. 

\section*{Acknowledgments}
The authors acknowledge the NSFC under grants U1938113 and 11773035. This work is supported by the Scholar Program of Beijing Academy of Science and Technology (DZ:BS202002) and CSST projects: CMS-CSST-2021-B07 and CMS-CSST-2021-B03.

\end{document}